\documentclass[11pt,a4paper]{article}

\newcommand{\swift}{{\it Swift}}
\newcommand{\chandra}{{\it Chandra}}
\newcommand{\xmm}{{\it XMM-Newton}}
\newcommand{\exosat}{{\it EXOSAT}}
\usepackage{natbib,graphicx}
\long\def\symbolfootnote[#1]#2{\begingroup%
\def\thefootnote{\fnsymbol{footnote}}\footnote[#1]{#2}\endgroup}

\begin{document}

\section*{The phases of X-ray emission of RS\,Oph}

\begin{center}
Jan-Uwe {\sc Ness}$^{1,}$\symbolfootnote[1]{\scriptsize Chandra Fellow at ASU;
Any opinions, findings, and conclusions or recommendations expressed in
this article are those of the author[s] and do not necessarily reflect the
views of SAO or NASA.
}
Sumner {\sc Starrfield}$^1$,
Kim L. {\sc Page}$^2$, Julian P. {\sc Osborne}$^2$,
Andy P. Beardmore$^2$, Jeremy J. {\sc Drake}$^3$\\
\vspace{.4cm}

{\small
$^1$School of Earth and Space Exploration, Arizona
State University, Tempe, AZ 85287-1404, USA\\
E-Mail: [Jan-Uwe.Ness,sumner.starrfield]@asu.edu\\
$^2$Department of Physics \& Astronomy, University of Leicester, Leicester, LE1 7RH, UK\\
$^3$Harvard-Smithsonian Center for Astrophysics, 60
Garden Street, Cambridge, MA 02138, USA
}
\end{center}

{\bf Abstract}
The recurrent symbiotic nova RS\,Oph reoccurred after 21 years on
12 February 2006. In contrast to the 1985 outburst, much denser
coverage with X-ray observations was achieved. \swift\ observed
RS Oph up to several times a day while \chandra\ and \xmm\
observed two to four times during each phase of evolution.
While the \swift\ observations provide high resolution in time,
the \chandra\ and \xmm\ observations provide high spectral
resolution. Refined models can be constrained by the grating
spectra, and interpolation of the model parameters can be
constrained by the wealth of \swift\ observations.
We compared the \swift\ light curve with six X-ray observations
taken with \exosat\ during the 1985 outburst. We found that
the decay from the supersoft X-ray binary (SSS) phase had
been observed.\vspace{-.2cm}

\section{Introduction}\vspace{-.2cm}

 RS\,Oph is a Symbiotic Recurrent Nova (RN) that recurs on time scales
of $\sim 20$\,years. The explosion occurs in a binary system
consisting of a white dwarf (WD) near the Chandrasekhar limit and
a red giant companion orbiting with a period of
$455.72\,\pm\,0.83$\,days \citep[see][]{dob94,sho96,fek00}.
On the surface of the WD a layer of hydrogen-rich
material is built up by accretion of material from the
red giant. When enough material is accreted,
a thermonuclear explosion occurs, and an optically thick shell
of ejected material initially blocks all high-energy radiation
produced by nuclear burning. The radiative energy output occurs
primarily in optical light until the shell clears. During this
early phase, all X-ray emission from RS\,Oph originates from a
shock produced by the expanding shell colliding with the
pre-existing wind of the red giant\citep{rsoph_iau1,bode06,sokoloski06}.
After the shell has cleared, X-ray emission
from the WD dominates the shock emission, and
X-ray observations allow a view deep into the outflow
\citep{rsoph_iau2,osborne06,ness_rsoph}. The X-ray spectrum during
this phase resembles that of the supersoft X-ray binary sources
\citep[SSS:][]{heuvel,kahab}. After nuclear burning on
the WD surface has turned off, residual X-ray emission from the
shock and from the ionized surrounding material is observed
\citep{rsoph_iau3}.\vspace{-.2cm}

\section{X-ray Observations}\vspace{-.2cm}

 The first X-ray observations of RS\,Oph were carried out
during the 1985 outburst with \exosat. In contrast, the 2006
outburst was monitored in X-rays by RXTE and \swift,
\citep{bode06,osborne06,sokoloski06}, while twelve observations were
carried out with \chandra\ and \xmm, using the high
spectral resolution of the grating spectrometers.

\begin{figure}[!ht]
\resizebox{\hsize}{!}{\includegraphics{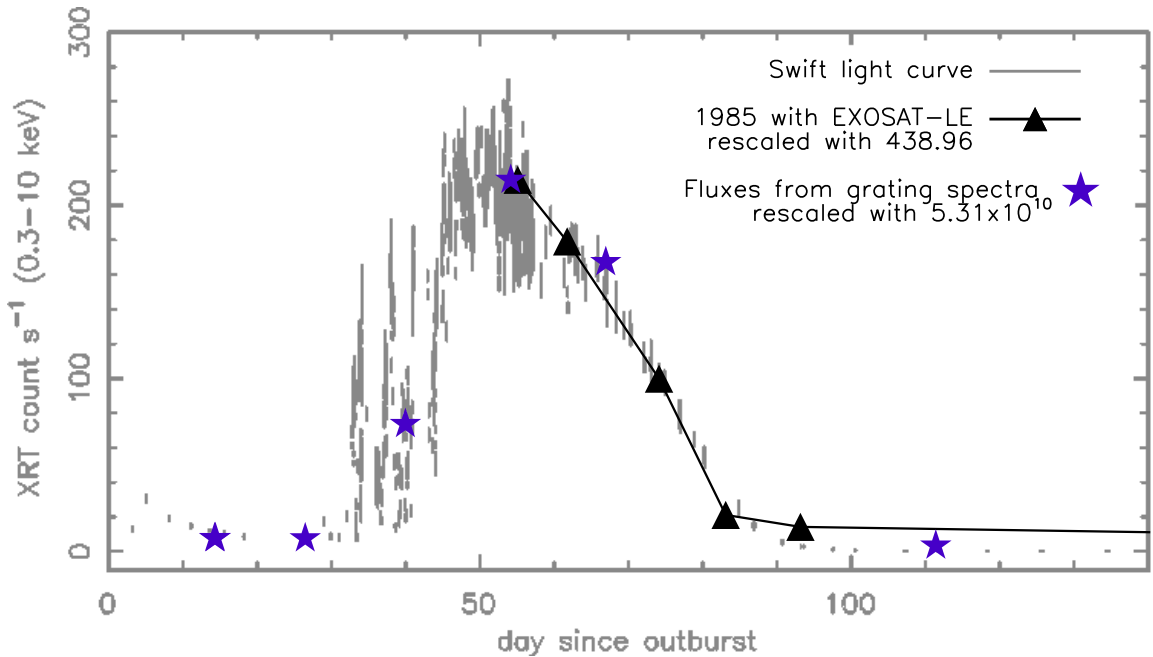}}
\caption{\label{lc1985}Light curve of the 2006 outburst taken
with \swift\ XRT \citep[grey;][]{osborne06} compared to \exosat\
(LE) count rates measured during the 1985 outburst
\citep[black;][]{mason87} and \chandra/\xmm\ grating
fluxes (blue; Table~\ref{tab1}). The count rates and fluxes
were rescaled with a constant factor (given in the legend)
to reproduce the XRT count rate on day 55.}
\end{figure}

 X-ray observations with \swift\ started three days
after outburst, and the evolution was followed until day
251 (2006, October 22). A total of 386.5\,ksec of exposure were
obtained, and in Fig.~\ref{lc1985} we show the X-ray
Telescope (XRT) light curve as presented by \cite{osborne06}. 
Until day $\sim 30$ an emission level between 10 and 30
counts per second was detected with a hard spectrum from the
shock \citep{bode06}. The ascent into
the SSS phase occurred very rapidly with an increase in count
rate by an order of magnitude, followed by strong variations in
brightness \citep{atel801}. The decay occurred after day 55, and
in Fig.~\ref{lc1985} we compare the Swift light curve with
observations of the 1985 outburst carried out with the Low Energy
(LE; 0.05-2 keV) Imaging Telescopes aboard \exosat\
\citep[black triangles;][]{mason87}.
We rescaled the \exosat\ count rates with the constant factor
given in the legend. With this factor the rescaled count rate
and the XRT count rate are the same on day 55.

In Table~\ref{tab1}
we summarize all grating observations of RS Oph taken in 2006
with observation date, day after outburst, instrumental
setup, observation ID, and net exposure time. We extracted
effective areas for each spectral bin from the calibration of
\chandra\ and \xmm\ using the respective analysis software
packages CIAO and SAS. Since the grating spectrometers have
very little photon redistribution,
the values of measured count rates in each spectral
bin can be converted to photon fluxes by division by the
effective areas. Energy fluxes are then obtained using the
wavelength of the corresponding grid points. This approach is
more accurate than defining a model over which to integrate
the fluxes. In the last column of Table~\ref{tab1} we list
the measured fluxes integrated over the wavelength range
7--38\,\AA\ (0.33--1.8\,keV). We include these values in
Fig.~\ref{lc1985}, rescaled in the same way as the \exosat\
count rates.\vspace{-.2cm}

\begin{table}[!ht]
\begin{flushleft}
\renewcommand{\arraystretch}{1.1}
\caption{\label{tab1}Grating observations of RS Oph in 2006}
\begin{tabular}{rp{.5cm}llrlr}
Date (start--stop)& day$^{[a]}$& Mission & Grating/detector & ObsID & net time (ksec)& flux$^b$\\
\hline
\ Febr.\,26,    15:20--18:46 & 13.8 & \chandra & HETG/ACIS & 7280 &  \multicolumn{2}{l}{9.9\hfill $9.0\,\pm\,0.20$}\\
\ Febr.\,26,    17:09--23:48 & 13.9 & {\it XMM} & \multicolumn{2}{l}{RGS\hfill 0410180101} & \multicolumn{2}{l}{23.8\hfill $9.1\,\pm\,0.05$}\\
\ Mar.\,10/11, 23:04--02:21 & 26.1 & {\it XMM} & \multicolumn{2}{l}{RGS\hfill 0410180201} &  \multicolumn{2}{l}{11.7\hfill $8.8\,\pm\,0.05$}\\
\ Mar.\,24,    12:25--15:38 & 39.7 & \chandra & LETG/HRC & 7296 &  \multicolumn{2}{l}{10.0\hfill $138\,\pm\,0.25$}\\
\ Apr.\,07/08, 21:05--02:20 & 54.0 & {\it XMM} & \multicolumn{2}{l}{RGS1/2\hfill 0410180301} &  \multicolumn{2}{l}{9.8/18.6\hfill $414\,\pm\,0.21$}\\
\ Apr.\,20,    17:24--20:28 & 66.9 & \chandra & LETG/HRC & 7297 &  \multicolumn{2}{l}{6.5\hfill $321\,\pm\,0.48$}\\
\ June\,04,     12:06--18:08 & 111.7 & \chandra & LETG & 7298 &  \multicolumn{2}{l}{19.9\hfill $0.62\,\pm\,0.008$}\\
\ Sept.\,06,    01:59--17:30 & 205.3 & {\it XMM} & \multicolumn{2}{l}{RGS\hfill 0410180401} &  \multicolumn{2}{l}{30.2\hfill $0.11\,\pm\,0.002$}\\
\ Sept.\,04,    10:43--22:26 & 203.6 & \chandra & LETG/HRC & 7390 &  \multicolumn{2}{l}{39.6\hfill $0.085\,\pm\,0.001$}\\
\ Sept.\,07,    02:37--14:29 & 206.3 & \chandra & LETG/HRC & 7389 &  \multicolumn{2}{l}{39.8\hfill $0.099\,\pm\,0.001$}\\
\ Sept.\,08,    17:58--23:36 & 207.9 & \chandra & LETG/HRC & 7403 &  \multicolumn{2}{l}{17.9\hfill $0.10\,\pm\,0.002$}\\
\ Oct.\,09/10,  23:38--13:18 & 239.2 & {\it XMM} & \multicolumn{2}{l}{RGS\hfill 0410180501} &  \multicolumn{2}{l}{48.7\hfill $0.07\,\pm\,0.001$}\\
\hline
\end{tabular}

$^{[a]}$after outburst (2006, Feb. 12.83)\hfill $^b$over 7--38\,\AA\ [10$^{-11}$\,erg\,cm$^{-2}$\,sec$^{-1}$]
\renewcommand{\arraystretch}{1}
\end{flushleft}
\end{table}

\section{Conclusions}\vspace{-.2cm}

The best coverage in X-rays was obtained with \swift, providing data
on an almost daily basis. The X-ray spectra obtained with \swift\ can 
be used to study the evolution of the temperature and the luminosity.
More detailed spectral information can be obtained from the \chandra\ and
\xmm\ spectra, however, more sophisticated models are required in order
to fully explore the grating spectra. For example, the SSS spectra need to
be modeled with atmosphere codes like PHOENIX \citep{petz05}.
 In a later step, the
model parameters obtained from the grating spectra can be interpolated,
using the \swift\ spectra as a constraint. In that way the 
interplay of both missions can achieve a full description of
the evolution of the outburst.

The \exosat\ measurements during the 1985 ourburst were interpreted
as residual shock emission by \cite{obrien92}, however it is now
clear that the LE observed the decline of the SSS phase. The comparison
between the XRT and the LE count rates reveals that the decay occurred
in exactly the same way. The spectrum was also very soft, and the
Medium Energy (ME; 1-50 keV) Proportional Counter has measured only
upper limits after day 80. \cite{obrien92} suspected that the SSS
emission may contaminate the shock emission, however, the XRT
and the \xmm\ and \chandra\ grating data show that the SSS spectrum
is far brighter compared to the shock emission at that stage than
had been expected.\vspace{-.2cm}

\section*{Acknowledgments}\vspace{-.2cm}
J.-U.N. gratefully acknowledges support provided by NASA through
\chandra\ Postdoctoral Fellowship grant PF5-60039 awarded by the
\chandra\ X-ray Center, which is operated by SAO for NASA under
contract NAS8-03060.
Any opinions, findings, and conclusions or recommendations expressed
in this article are those of the authors and do not necessarily
reflect the views of SAO or NASA.
S.S. received partial support from NSF and NASA grants to ASU.
K.L.P., J.P.O. acknowledge support by the Particle Physics and
Astronomy Research Council.

\small
\vspace{-.2cm}
\bibliographystyle{apj}
\bibliography{rsoph,astron,cn,jn}

\end{document}